UDC 577

# Epigenetics: What it is about?

E. Saade[1], V. V. Ogryzko[2]

[1]Faculty of Public Health, Lebanese University,
P.O. Box 6573/14 Badaro, Museum, Beirut, Lebanon

[2]CNRS UMR8126, Paris-Sud University, Gustave Roussy Institute
114, rue Edouard Vaillant, 94805 Villejuif Cedex-France

vogryzko@gmail.com

*Epigenetics has captured the attention of scientists in the past decades, yet its scope has been continuously changing. In this paper, we give an overview on how and why its definition has evolved and suggest several clarification on the concepts used in this field. Waddington coined the term in 1942 to describe genes interaction with each other and with their environment and insisted on dissociating these events from development. Then, Holliday and others argued that epigenetic phenomena are characterized by their heritability. However, differentiated cells can maintain their phenotypes for decades without undergoing division, which points out the limitation of the «heritability» criterion for a particular phenomenon to qualify as epigenetic. «Epigenetic stability» encompasses traits preservation in both dividing and non dividing cells. Likewise, the use of the term «epigenetic regulation» has been misleading as it overlaps with «regulation of gene expression», whereas «epigenetic information» clearly distinguishes epigenetic from genetic phenomena. Consequently, how could epigenetic information be transmitted and perpetuated? The term «epigenetic templating» has been proposed to refer to a general mechanism of perpetuation of epigenetic information that is based on the preferential activity of enzymes that deposit a particular epigenetic mark on macromolecular complexes already containing the same mark. Another issue that we address is the role of epigenetic information. Not only it is important in allowing alternative interpretations of genetic information, but it appears to be important in protecting the genome, as can be illustrated by bacterial endonucleases that targets non methylated DNA – i. e. foreign DNA – and not the endogenous methylated DNA.*

**Does epigenetics contradict the Central Dogma of Molecular Biology**? Epigenetic phenomena represent a topic of active research in current biology. This field of study is popular in part because it goes «against the grain» of the Central Dogma of Molecular Biology. If understood in an oversimplified way, the Central Dogma professes a «genocentric» view of a biological system, according to which all information necessary to define the state of an organism (up to the reaction norm due to the effects of environment) is contained in the sequence of its genome. While more or less adequate when one describes simple single-cellular organisms, such as bacteria, the «genocentric» view runs into problem when one start to deal with the multicellular organisms that exhibit the phenomenon of cellular differentiation. Different cell types from the same organism can be taken from an organism and propagated in a cell culture, in identical environmental conditions. The genome sequence of these different cell types is identical (as demonstrated by the phenomenon of somatic cloning and iPS cell reprogramming [1, 2]), yet they exhibit very different and stable phenotypes. This observation suggests that there should be additional information responsible for the maintenance of the stable differentiated phenotypes. Understanding the nature and the mechanisms of processing and propagation of this information represents one of the major topics in the epigenetic studies.

In fact, the Central Dogma, as it was first formulated by Francis Crick in 1958 and re-stated in a Nature paper published in 1970 [3] asserts: «The central dogma of molecular biology deals with the detailed residue-by-residue transfer of sequential information. It states that such information cannot be transferred from protein to either protein or nucleic acid». A careful reading of this definition shows that it leaves the door open for other types of information that could be required to specify the state of the organism (*e. g.* stored in macromolecular conformations, interactions, post-translational modifications and alternative states of genetic networks) and might propagate independently from the DNA sequence. Thus, despite a conceptual tension between the «genocentric» view of biological systems and the notion of epigenetic phenomena, the latter is compatible with the Central Dogma of Molecular Biology as it was formulated by Crick.

**Why the definition of epigenetics changed?** – *Natura abhorret vacuum*[1]. The understanding of what constitutes the subject of epigenetics underwent changes from the time of its conception. The term «epigenetic» was coined by Conrad Waddington in 1942 referring to mechanisms in which genes within an organism interact with each other and their environment to create a phenotype (the related notion of «epigenesis» dates back to Aristotle (On the Generation of Animals)). [4]. One of the main metaphors in this early version of epigenetics was one of «epigenetic landscape» that defines different stable trajectories that a cell can take during development and differentiation.

In this picture, genes are responsible for generating various epigenetic landscapes, which still would be consistent with the «genocentric» view on organisms. However, the main focus of Waddington's interest was on *relative decoupling of development from genes*. A canonical example of such decoupling was the phenomenon of phenocopy [5], *i. e.*, a developmental abnormality that is phenotypically identical to that caused by genetic mutations, but induced by factors that do not change genotype, such as heat shock [6]. This phenomenon, together with the reciprocal observation that not every change in genotype is reflected in phenotype, constituted the main argument against the simple notion of development completely determined by genetic information.

However, nowadays the term «epigenetics» is used in a different meaning (due originally to Holliday) and emphasizes «*heritability*» as a necessary requirement for a phenomenon to qualify as epigenetic. The first ex-perimental models for this redefined field of research were DNA methylation in [7], position effect variegation (PEV) in *Drosophila melanogaster* [8] and silencing phenomenon in yeast [9].

How this change in the meaning of the term «epigenetic» can be explained? Most probably it came about as the result of the shift in the meaning of the term «genetic».

As originally introduced, Genetics is «science of inheritance» [10]. If one strictly abides by this terminology, epigenetics cannot study heritable variations at all, simply because by definition they all should be considered as genetic.

However, after the solving of DNA structure by Watson and Crick and consequent deciphering of genetic code, genetic information became justifiably identified with a particular sequence of nucleotides on DNA or RNA. Accordingly, the meaning of the term «genetic» shifted, as all things «genetic» became firmly associated with the sequences of DNA or RNA.

On another hand, the phenomenon of stability of differentiated phenotypes (and related phenomena, such PEV and silencing) demonstrates existence and importance of heritable variations that cannot be accounted for by genetic variations. To fill the conceptual void due to the shift in the meaning of the term «genetic», the term «epigenetic» changed its meaning as well, motivated by the need to describe the class of heritable phenomena not encoded in DNA (/RNA) sequence and thus left outside of the scope of newly redefined genetics.

**Is heritability essential? Epigenetic stability and epigenetic information**. The above discussion suggests an explanation of how the modern meaning of the term «epigenetic» came about. However, it appears that the requirement of «heritability», currently emphasized as the criterion for a particular phenomenon to fall under the scope of epigenetic research, might be too restrictive.

Terminally differentiated cells, such as neurons and muscle cells, live for decades maintaining their distinct phenotypic differences in spite of environmental stresses, thermal noise and DNA damage/repair [11, 12]. These cells do not proliferate, thus the term «heritable» does not apply to their distinct and stable traits. On the other hand, it is reasonable to expect that maintenance and propagation of information responsible for these traits employs mechanisms that are similar to those that are utilized in the replicating cells (*e. g.*, DNA methylation, histone modifications, *etc*.). Thus, although one cannot use the criteria of «heritability», it makes good sense to qualify the variations between different cell lineages (and the underlying mechanisms) as epigenetic. The term «epigenetic stability» refers to a broader phenomenon that encompasses maintenance of phenotypic traits in both nonreplicating and replicating cells independent from DNA sequence [13, 14].

---

[1]*Francois Rabelais «Gargantua and Pantagruel'», 1530s.*

On the other extreme, some authors suggest to understand epigenetic phenomena in a very broad sense, defining them as any kind of inherited traits that do not exhibit Mendelian behavior [15]. This definition appears to be too loose, as it will consider as epigenetic all bacterial, organelle and viral genetics, which is based on the primary structure of nucleic acid, while does not conform to the rules of Mendel. It is therefore advisable to restrict the use of term epigenetic to the information that is not contained in nucleic acid sequence.

Therefore, in what follows, epigenetic information will be defined as the information that is required in order to specify the state of an organism in addition to genetic information (nucleotide sequence) and reaction norm. The mechanisms of processing, storage and transmission/propagation of epigenetic information are the subject of molecular epigenetics, a relatively new field of research.

**Is the notion of «epigenetic regulation» useful or confusing**? The value of a scientific concept depends on: (*i*) how well it can capture a particular class of phenomena by clearly distinguishing it from other phenomena, (*ii*) whether it can stimulate new directions of research and (*iii*) whether it allows one to convey complicated ideas in a succinct and lucid fashion.

Unlike the term «epigenetic information», another widely used notion – that of «epigenetic regulation» (or epigenetic control) – does not meet these criteria.

The major problem with this notion stems from the difficulty in clearly defining its scope. Its original intention was to describe regulation of gene expression during development and differentiation of multicellular organisms (in line with the Waddington's definition of epigenetics). After the discovery of parallels between PEV in drosophila and silencing phenomena in yeast, as well as with finding of epigenetic-like phenomena in other microorganisms [16, 17], it became evident that the studies of epigenetic mechanisms cannot be limited to multicellular organisms and their development. However, broadening of the scope of «epigenetic regulation» complicates its demarcation from the general notion of «regulation of gene expression», long used in molecular biology and genetics. Genetic circuits and networks had been studied (not necessarily in the context of development) for decades in the field of molecular genetics [18]. Genetic interactions (*e. g.*, epistasis) have been the subject of evolutionary genetics from its very beginning (with the studies of evolution of dominance being a prominent example [19]). It appears that the term «epigenetic regulation» currently serves merely to replace «regulation of gene expression», without any added value in the transaction. The introduction of new molecular players, such as noncoding RNA, to the list of epigenetic mechanisms adds nothing but confusion.

A telling illustration of how the term «epigenetic regulation» is encroaching on the turf of classical molecular genetics is a name changing game that occurred in the field of bacteriophage lambda. The transition between lyzogenic and lytic lifecycles of this phage was originally termed «genetic switch» [20], serving as an early paradigm for regulation of gene expression. However, currently the same phenomenon is often referred to as «epigenetic switch» [21], also used by Ptashne himself [22], including a mock change of his classic book's cover in a recent talk at Pasteur, Paris (EMBO Workshop on the Operon Model and its impact on modern molecular biology, 17–20 May 2011). Note that, whereas the notions of genetic and epigenetic information clearly describe distinct aspects of this system (*i. e.*, sequence of nucleotides and alternative states of gene activity, respectively), the notions of genetic and epigenetic switch (or else control, regulation, or a circuit) would have exactly the same meaning.

Given that chromatin is considered a principal carrier of epigenetic information, one way to remedy the «encroaching problem» would be to limit the scope of epigenetic regulation to chromatin-based regulation. However, such clarification does not seem wise, as the fields of chromatin and epigenetics are only overlapping, not identical. Not every chromatin modification (histone PTM, DNA methylation, replacement histone) is heritable (or stable) and thus could serve as an epigenetic mark, and not every epigenetic change is encoded in chromatin state (as illustrated by such trans-acting epi-genetic factors as alternative states of genetic networks, prions, structural/cortical inheritance [16, 18, 23, 24]).

In contrast to «epigenetic regulation», the scope of «epigenetic information» is, by construction, clearly defined – namely, as something additional to (and hence, different from) «genetic information». Moreover, after distinguishing between the two notions, the time tested conceptual framework established to study genetic information can be now transferred to the newer field of epigenetics, immediately suggesting plethora of questions and directions of research. How epigenetic information is reproduced and transmitted? How is it recorded *de novo* and how it is read and/or erased? Does the notion of «epigenetic information damage» make any sense? If so, can it be repaired and/or are there mechanisms of epigenetic damage response, such as specialized cell cycle checkpoints [12]?

In addition to the clearly defined scope, the value of the term «epigenetic information» is evident from the new questions it opens and the avenues of research it stimulates.

**Epigenetic templating – «*like draws to like*».** How epigenetic information can be propagated and maintained? Studies of DNA methylation (an earliest recognized epigenetic mark) provided an early clue on one mechanism, based on a different behavior of the so called maintenance methylase enzyme towards un-methylated versus semi-methylated DNA (Fig. 1). The semiconservative DNA replication of a methylated double strand gives rise to semimethylated bases comprising a parental strand and a newly synthesized one. Methyltransferase enzymes bind to the semimethylated sites where they methylate the new strand. The modifying machinery is not recruited to un-methylated strands.

Not surprisingly, this mechanism differs from that of replication of genetic information. Instead of Watson-Crick base complementarity, it is rather based on the ancient principle «*like draws to like*». Later on, Francis Crick proposed a similar mechanism as the molecular basis for neurobiological memory, essentially an epigenetic phenomenon [25]. He postulated that (*i*) the strength of a synapse is determined by phosphorylation of a protein molecule, (*ii*) this protein can form dimers (or oligomers) and (*iii*) the kinase responsible for the modification will only modify monomer in a dimer that has second monomer already phosphorylated.

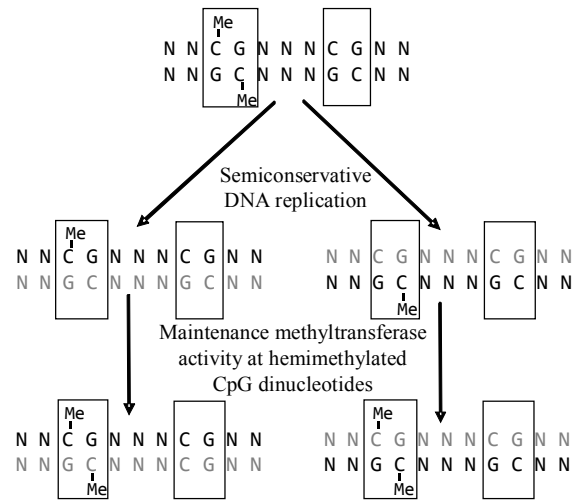

Fig. 1. Methylation maintenance. After replication, only semimethylated bases are converted into fully methylated

Further one, consistent with the recognized role of chromatin as a principal carrier of epigenetic information, very similar mechanisms have been suggested for perpetuation of post-translational histone modifications. Some histone acetyl transferase (HAT) complexes could be preferentially recruited to acetylated chromatin due to the presence of bromodomains, recognizing acetylated lysines, whereas the chromodomain containing histone methyl transferase (HMT) complexes could be likewise recruited to methylated chromatin [26–28]. Also, domain structure of some protein kinases suggests that similar mechanism could operate not only on the level of chromatin, but in controlling the organization of cytoplasm as well. Tyrosine kinases often contain SH2 domains, which recognize phosphotyrosine, whereas some serine/threonine kinases contain FHA domains that recognize phosphoserine/phosphothreonine. If some of the targets of these kinases form dimers (or oligomers), their phosphorylation status can be perpetuated according to the Crick's original proposal.

To emphasize the difference of this way to propagate information from the canonical Watson-Crick base-base complementarity mechanism, Vasily Ogryzko's group used the term «epigenetic templating» [13, 14] to refer to a general mechanism of perpetuation of epigenetic information that is based on the preferential activity of enzymes that deposit a particular epigenetic mark on macromolecular complexes already containing the same mark (Fig. 2). They tested whether this model can also apply to variant histones, putative epigenetic marks on chromatin different from post-translational histone modifications.

**What are the roles of epigenetic information and their evolutionary relations?** Why Life needs epigenetic information? As was implied above, one of its roles is in facilitating maintenance of differentiated cell phenotypes during development of multicellular organisms, that is, in supporting *alternative interpretations of genetic information*. However, one might argue that from evolutionary perspective, a more fundamental and primary role could have been *protection of genetic information*.

An important aspect of functioning of genetic information is its formal («mechanical») character. The replication, transcription, recombination and other enzymatic machineries are designed to treat all nucleotide sequences equally, regardless of their meaning, *i. e.*, of what do they encode. However, this «equal treatment» allows viruses and other parasitic genetic elements to take advantage of the cellular resources and propagate their own genetic information. In order to protect themselves, cells have developed ways to label their own genetic information so that it could be recognized as their own and distinguished from the foreign one. One can consider an example of restriction-methylation system, a defense mechanism developed by bacteria. This system contains a restriction endonuclease and a methylase enzymes that can target the same DNA sequence, depending on its methylated status (Fig. 2). A foreign DNA invading the cell is not methylated and thus becomes a target of the endonuclease. However, the same target sequence present originally in the cell is fully methylated, and the methylated state is maintained after replication due to the action of maintenance methylase (as described in the previous section). Thus, cellular machinery that helps it to protect its own genetic information is based on *recognition of epigenetic information* associated with DNA.

Moreover, the same mechanism of replication of methylated status of DNA serves additional protective purpose, by helping bacteria to *prevent errors* during replication of genetic information. In order to correctly remove the erroneously incorporated nucleotide, mismatch repair mechanisms need to recognize the parental strand of the newly duplicated DNA. While the newly replicated DNA is in the semimethylated state, the parental strand is labeled with methyl, which allows

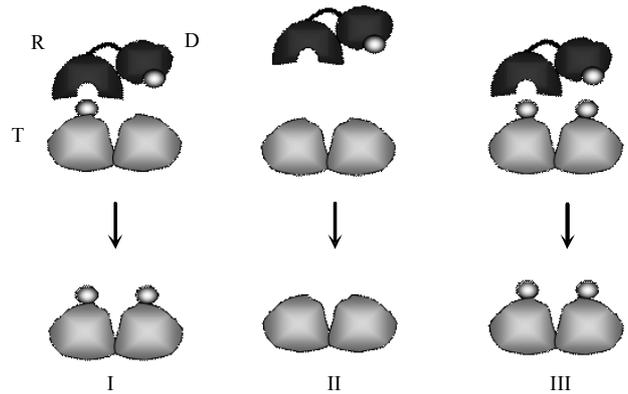

Fig. 2. Scheme of epigenetic templating [14]. In the case I, the modifying machinery is recruited to sites exhibiting a particular mark and therefore it amplifies the information. In the case II, if the mark is absent, the machinery is not recruited. In III, the recruitment occurs but no additional modification takes place because target sites are already modified

the mismatch repair machinery to correctly identify which base of a mismatch pair has to be removed.

The mismatch repair mechanism illustrates how the notion of «epigenetic information» allows one to convey a complicated idea in an economical fashion, demonstrating an additional value of this concept. Indeed, the discrimination between the parental and daughter strands can work only if there is a time window when the semimethylated DNA has not yet been converted to the fully methylated state. This feature can be formulated in the following elegant way: the increased accuracy of replication of genetic information in bacteria is ensured by *the difference between the rates of replication of genetic and epigenetic information*. Incidentally, this makes mismatch repair an example of a general kinetic proofreading scheme [29].

In the above examples, epigenetic information acts as a part of protection system serving to differentiate the «self» from the «other» and the «old» from the «new». Thus, at least two distinct roles of epigenetic information can be distinguished: interpretation of genetic information and its protection. What could be the evolutionary relation between these two functions? It seems reasonable to propose that the protection of genetic information had emerged first in evolution, since it would be beneficial already for the single cellular organisms, which typically have much less need for cell differentiation. Only later, epigenetic mechanisms could be recruited to play a role in cell differentiation. Supporting this idea is the evidence that many epigenetic mechanisms appear to be related to mechanisms of suppression of parasitic genetic elements, as illustrated by RNA-directed DNA methylation [30, 31], the tendency of eukaryotic cells to heterochromatinize tandem repeats, in either RNA dependent or independent way [32, 33], *etc*. This observation allows us to suggest that the epigenetic machinery was established by Life for «genome protection» purposes first. In the further course of evolution, however, the mechanisms of processing of epigenetic information – which allow recognition of different epigenetic marks and channeling the signals encoded in these marks along appropriate response pathways – were recruited for other purposes, such as for stabilizing different alternative states of the same organism.

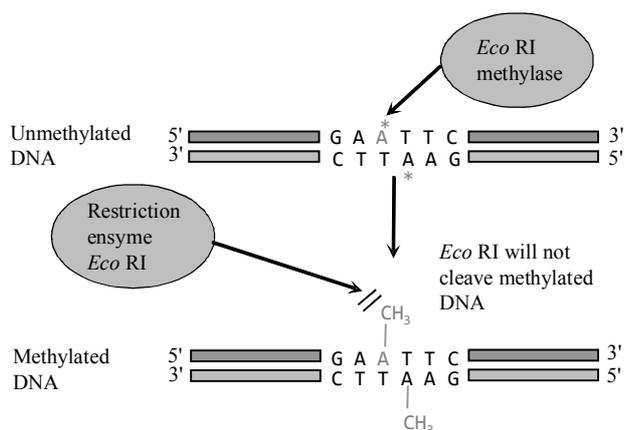

Fig. 3. Restriction endonucleases are a natural part of the bacterial defense system

**Conclusion**. With the ongoing progress in chromatin research, the development of iPS cell technologies and the emergence of evo-devo paradigm in evolutionary studies, research in epigenetics remains on the forefront of modern biology. As it happens with fast-developing fields, the scope of epigenetics tends to widen. An increasing number of molecular, cell and evolutionary biologists become motivated to position their research as epigenetic, both to keep up with the fashion and to have a better chance in funding or publishing in trendy journals. The many various views on what constitutes the proper scope of epigenetics call for development of a unified conceptual framework for this field. We hope that the proposed clarifications of the notions of «epigenetic stability», «epigenetic information» and «epigenetic templating», as well as our discussion of changing roles of epigenetic mechanisms in evolution will contribute into this worthy endeavor.

**Funding**. This work was supported by «La Ligue Contre le Cancer» (9ADO1217/1B1-BIOCE), the «Institut National du Cancer» (247343/1B1-BIOCE) and ARC foundation (SFI20121205936).

**Acknowledgements**. The authors thank Drs. Marc Lipinski, Murat Saparbaev and Alexander Ishchenko for many fruitful discussions.

Що ж таке епігенетика?

Е. Саад, В. В. Огризко

Резюме

*Епігенетика привертає увагу вчених уже декілька десятиліть, але значення вкладеного в неї сенсу постійно змінюється. Ми обговорюємо причини таких змін і пропонуємо декілька пояснювальних коментарів. Уоддінгтон увів термін «епігенетика» в 1942 році для описання взаємодії генів, важливих для розвитку організму, між собою та з навколишнім середовищем. Надалі Холлідей та інші наполягали на понятті «успадковуваності» як необхідній характеристиці епігенетичних явищ. Однак диференційовані клітини зберігають свої фенотипи протягом десятиліть і при цьому не діляться, що вказує на обмеженість «успадковуваності» як критерію того, щоб певне явище розглядати як епігенетичне. «Епігенетична стабільність» запропонована як більш загальне поняття, яке означає збереження характеристик в обох клітинах: які ділять і не діляться. З іншого боку, термін «епігенетична регуляція» призводить до плутанини, оскільки його значення суттєво перекривається або навіть повністю збігається за змістом з виразом «регуляція експресії генів», тоді як «епігенетична інформація» чітко розмежовує епігенетичні і генетичні явища. І постає питання, яким чином епігенетична інформація може відтворюватися? Ми запропонували термін «epigenetic templating» для визначення механізму відтворення епігенетичної інформації, заснованого на тому, що ферменті, які ставлять певну епігенетичну мітку, віддають перевагу макромолекулярним субстратам, які вже містять подібну мітку. Нарешті ми торкаємось питання щодо ролі епігенетичної інформації. Вона потрібна не лише для альтернативних інтерпретацій генетичної інформації, але й для захисту геному, як це проілюстровано нами на прикладі бактерійних ендонуклеаз, які атакують неметильовану (тобто чужорідну) ДНК і не пошкоджують метильованої (тобто власної) ДНК.*

*Ключові слова*: хроматин, спадковість, центральна догма, метилювання ДНК, генетичний перемикач, еволюція.

Что же такое эпигенетика?

Э. Саад, В. В. Огрызко

Резюме

*Эпигенетика привлекает внимание ученых уже несколько десятилетий, но значение вкладываемого в нее смысла постоянно меняется. Мы обсуждаем причины таких изменений и предлагаем несколько поясняющих комментариев. Уоддингтон ввел термин «эпигенетика» в 1942 году для описания взаимодействия генов, важных для развития организма, друг с другом и с окружающей средой. Позже Холлидей и другие настаивали на понятии «наследуемости» как необходимой характеристике эпигенетических явле-*

ний. Однако дифференцированные клетки сохраняют свои фенотипы на протяжении десятилетий и при этом не делятся, что указывает на ограниченность «наследуемости» как критерия того, чтобы определенное явление рассматривать как эпигенетическое. «Эпигенетическая стабильность» предложена как более общее понятие, означающее сохранение характеристик в обеих делящихся и не делящихся клетках. С другой стороны, термин «эпигенетическая регуляция» ведет к путанице, поскольку его значение сильно перекрывается или даже совпадает по смыслу с выражением «регуляция экспрессии генов», в то время как «эпигенетическая информация» четко разграничивает эпигенетические и генетические явления. Далее следует вопрос, каким образом эпигенетическая информация может воспроизводиться? Мы предложили термин «epigenetic templating» для обозначения механизма воспроизведения эпигенетической информации, основанного на том, что ферменты, ставящие определенную эпигенетическую метку, предпочитают макромолекулярные субстраты, уже содержащие подобную метку. Наконец, мы касаемся вопроса о роли эпигенетической информации. Она нужна не только для альтернативных интерпретаций генетической информации, но и для защиты генома, как это проиллюстрировано нами на примере бактериальных эндонуклеаз, атакующих неметилированную (то есть чужеродную) ДНК и не повреждающих метилированной (то есть своей) ДНК.